\documentclass[10pt,letterpaper]{article}
\usepackage{amsmath}
\usepackage{amsfonts}
\usepackage{amssymb}
\usepackage[dvips]{epsfig}

 \advance \textwidth by 0.1cm
 \advance \textheight by 0.06cm
\def\##1{{\bf #1}}
\def\=#1{\underline{\underline{#1}}}

\def\+#1{\underline{\bf #1}}
\def\*#1{\underline{\underline{\bf #1}}}

\def\les{\left[}
\def\ris{\right]}
\def\lec{\left\{}
\def\ric{\right\}}

\def\.{\mbox{ \tiny{$^\bullet$} }}

\def\eps{\epsilon}
\def\epso{\epsilon_{\scriptscriptstyle 0}}

\def\muo{\mu_{\scriptscriptstyle 0}}

\def\epsr{\epsilon_r^{\prime}}
\def\epsi{\epsilon_r^{\prime\prime}}

\def\nplus{n_+}
\def\nminus{n_-}

\def\om{(\omega)}
\def\Om{(\Omega)}

\def\ux{\hat{\#x}}
\def\uy{\hat{\#y}}
\def\uz{\hat{\#z}}

\def\pa{\frac{p\pi}{a}}
\def\cpa{\cos\left(\pa x\right)}
\def\spa{\sin\left(\pa x\right)}
\def\qb{\frac{q\pi}{b}}
\def\cqb{\cos\left(\qb y\right)}
\def\sqb{\sin\left(\qb y\right)}

\begin{document}

\begin{center}

{\bf When does the choice of the refractive index of a linear, homogeneous, isotropic, active, dielectric medium matter?}\\

\bigskip

Akhlesh Lakhtakia,$^1$ Joseph B. Geddes III,$^{2\ast}$ and Tom G. Mackay$^{3}$

\bigskip\bigskip
$^1$Department of Engineering Science \& Mechanics,\\
 Pennsylvania State University, University Park, PA 16802 \\
$^2$Beckman Institute for Advanced Science and Technology, \\
University of Illinois at Urbana-Champaign,
Urbana, IL 61801\\
$^3$School of Mathematics,  University of Edinburgh, Edinburgh EH9 3JZ, United Kingdom\\
$^\ast$Corresponding author: {geddes@uiuc.edu}  

\end{center}


\begin{abstract}
Two choices are possible for the refractive index of a linear,
homogeneous, isotropic, active, dielectric material. Either
of the choices is adequate for obtaining
 frequency--domain solutions for (i)  scattering   
by slabs, spheres, and other objects of
bounded extent; (ii) guided--wave propagation in  homogeneously filled,
cross--sectionally uniform, straight
waveguide sections with perfectly conducting walls; and 
(iii) image form\-ation due to flat lenses. The correct choice does
matter for the half--space problem, but that problem is not realistic.
\end{abstract}

\section{Introduction}

A century and a half after the unification of light with electricity and magnetism, certain aspects of electromagnetic wave propagation in a linear, homogeneous, isotropic, dielectric material continuum remain unsettled. One of those aspects is as follows.  If $\eps_r\om=\epsr\om+i\epsi\om$ denotes the relative
permittivity~---~where $\epsr\in\mathbb{R}$, $\epsi\in\mathbb{R}$,
and an $\exp(-i\omega t)$ time--dependence
is implicit~---~then the refractive index $n\om$ satisfies the dispersion relation $n^2\om
=\eps_r\om$.
 Is the refractive index equal to $\nplus$ or $\nminus=-\nplus$, where
$\mbox{Im}( {\nplus})>0$?

Suppose that a material is passive at a specific frequency $\Omega$, i.e., $\epsi\Om>0$. The general consensus is that the
refractive index is equal to $\nplus\Om$.
If the material is active at the frequency $\Omega$ (i.e., $\epsi\Om<0$), however, disagreements  on the correct choice of the refractive index have surfaced. We are exclusively concerned with linear, homogeneous, isotropic, active, dielectric materials in this communication, subject to the additional stipulations that the constitutive properties are spatially local and unalterable by the passage of an electromagnetic signal.

Many reasons have been provided for the refractive index of such a material to be
$\nminus\Om$~\cite{Silverman1993}--\cite{WX2007} and $\nplus\Om$~\cite{CFW2006}--\cite{NC2007e}. A direct time--domain solution of the Maxwell equations has recently upheld the possibility of $\nplus\Om$ as the refractive index~\cite{GML2007}. Even more importantly, the adoption of the time--domain procedure implies that the refractive index cannot be determined at a single frequency by frequency--domain analyses, if the material is active at that frequency. By virtue of this argument, it is possible for two different active materials to have the same permittivity but opposite refractive indexes at a certain frequency~\cite{NS2007}.

Thus, in order to choose the refractive index of an active material at a specific frequency, the solution of a time--domain problem is needed with an incident signal of sufficient bandwidth to cover the frequency of interest (though it appears that a frequency--domain
algorithm may sometimes suffice to determine the variation of
$n\om$ vs. $\omega$ over
a sufficiently wide $\omega$--range~\cite{NS2007}--\cite{SkaarPRE}). The time--domain problem should be such that the correct choice of the refractive index in the corresponding frequency--domain problem must be consequential~---~e.g., the
reflection of a signal by a half--space filled with a material continuum~\cite{GML2007}. This naturally leads to the question: when is knowledge of the correct choice of the refractive
index essential for the solution of a frequency--domain problem?

In the following section, several commonly tackled frequency--domain boundary--value
problems are discussed, in order to answer the foregoing question. The selected
problems include scattering by objects of
bounded extent; guided--wave propagation in homogeneously filled straight
waveguide sections with perfectly conducting walls and uniform cross--section; and lensing by slabs.

\section{Frequency--domain Boundary--value Problems} \label{S2}

\subsection{Planewave response of a slab}\label{slab}
Let us begin by considering the electromagnetic  response of the region $0<z<L$ filled with an active material. The regions $z<0$ and $z>L$ are vacuous.

Because the dyadic Green function for free space can be expanded in terms of an angular spectrum of plane waves, we can consider the response of the slab to a plane wave without significant loss of generality. An incident plane wave can be represented by
\begin{equation}
{\#E}_{inc}(\#r,\omega)= \les a_s\, \uy + a_p \left(-\frac{\alpha_0}{k_0}\,\ux+\frac{\kappa}{k_0}\,\uz\right)\ris\,\exp\les i\left(\kappa x + \alpha_0 z\right)\ris\,,\quad z\leq 0\,,
\end{equation}
where $a_s$ and $a_p$ are the amplitudes of the $s$-- and the $p$--polarized components, respectively; $k_0$ is the free--space wavenumber; and
$\alpha_0 =+\sqrt{k_0^2-\kappa^2}$ with $\kappa\in\mathbb{R}$. The reflected and the transmitted plane waves are represented by
\begin{equation}
{\#E}_{ref}(\#r,\omega)= \les r_{s}\, \uy + r_{p}\left(\frac{\alpha_0}{k_0}\,\ux+\frac{\kappa}{k_0}\,\uz\right)\ris\,\exp\les i\left(\kappa x - \alpha_0 z\right)\ris\,,\quad z\leq 0\,,
\end{equation}
and
\begin{equation}
{\#E}_{tr}(\#r,\omega)= \les  t_{s} \,\uy +t_{p}\left(-\frac{\alpha_0}{k_0}\,\ux+\frac{\kappa}{k_0}\,\uz\right)\ris\,\exp\lec i\les\kappa x + \alpha_0 (z-L) \ris\ric\,,\quad z\geq L\,,
\end{equation}
respectively, while the electric field phasor inside the slab region is
\begin{eqnarray}
\nonumber
&&
{\#E}_{int}(\#r,\omega)= \les c_s\, \uy + c_p \left(-\frac{\alpha}{k_0 n}\,\ux+\frac{\kappa}{k_0 n}\,\uz\right)\ris\,\exp\les i\left(\kappa x + \alpha z\right)\ris \\
&&\qquad+\,
 \les d_s\, \uy + d_p \left(\frac{\alpha}{k_0 n}\,\ux+\frac{\kappa}{k_0 n}\,\uz\right)\ris\,\exp\les i\left(\kappa x - \alpha z\right)\ris\,,\quad 0 \leq z \leq L\,.
\end{eqnarray}
where $\alpha = +\sqrt{k_0^2n^2-\kappa^2}$. Standard techniques yield the amplitudes
\begin{equation}
\left.
\begin{array}{ll}
r_p = -a_p (e^{2i\alpha L}-1) (\alpha^2-n^4\alpha_0^2)/\Delta_p\,, &
t_p=-4n^2a_p e^{2i\alpha L}\alpha\alpha_0/\Delta_p\\[5pt]
c_p=-2a_pn\alpha_0(\alpha + n^2\alpha_0)/\Delta_p \,, &
d_p=-2a_pn\alpha_0(\alpha - n^2\alpha_0)/\Delta_p\\[5pt]
r_s = -a_s (e^{2i\alpha L}-1) (\alpha^2-\alpha_0^2)/\Delta_s\,,&
t_s=-4a_s e^{2i\alpha L}\alpha\alpha_0/\Delta_s\\[5pt]
c_s=-2a_s\alpha_0(\alpha + \alpha_0)/\Delta_s \,, &
d_s=2a_s\alpha_0(\alpha - \alpha_0)/\Delta_s
\end{array}\right\}\,,
\end{equation}
where
\begin{equation}
\left.
\begin{array}{ll}
\Delta_p= (e^{2i\alpha L}-1)(\alpha^2+n^4\alpha_0^2) -2 n^2(e^{2i\alpha L}+1)\alpha\alpha_0\\[5pt]
\Delta_s= (e^{2i\alpha L}-1)(\alpha^2+ \alpha_0^2) -2  (e^{2i\alpha L}+1)\alpha\alpha_0
\end{array}\right\}\,.
\end{equation}

The reflected, transmitted, and the internal electric field phasors are not affected by the change $n\rightarrow -n$; neither are the corresponding magnetic field phasors. Therefore, the fields \emph{everywhere} are unaffected whether $n=\nplus$ or $n=\nminus$ is chosen. The same conclusion is obtained if the plane $z=L$ is assumed to be perfectly conducting.

\subsection{Planewave response of a sphere}\label{sphere}
Let us next consider the spherical region $r<a$ occupied by an active
material, whereas the region
$r>a$ is vacuous. Without loss of generality, we take the sphere to be illuminated
by a linearly polarized plane wave  traveling along the $+z$ axis.
As is commonplace, the incident plane wave
is represented in terms of vector spherical harmonics,
$\#M_{\sigma m\nu}^{(j)}(\#w)$ and $\#N_{\sigma m\nu}^{(j)}(\#w)$~\cite{BH}, as
\begin{equation}
\label{eq7}
\#E_{inc}(\#r,\omega) = A \sum_{\nu=1}^\infty\,i^\nu\frac{2\nu+1}{\nu(\nu+1)}\,
\left[\#M_{o1\nu}^{(1)}(k_0\#r) -i\#N_{e1\nu}^{(1)}(k_0\#r)\right]\,,
\end{equation}
where $A$ is the amplitude.
The scattered and the internal electric  field phasors are also expressible in terms of vector spherical harmonics; thus~\cite{BH},
\begin{equation}
\#E_{sc}(\#r,\omega) =A \sum_{\nu=1}^\infty\,i^\nu\frac{2\nu+1}{\nu(\nu+1)}\,
\left[ia_\nu\#N_{e1\nu}^{(3)}(k_0\#r) -b_\nu\#M_{o1\nu}^{(3)}(k_0\#r)\right]\,,
\quad r \geq a\,
\end{equation}
and
\begin{equation}
\#E_{int}(\#r,\omega) =-A \sum_{\nu=1}^\infty\,i^\nu\frac{2\nu+1}{\nu(\nu+1)}\,
\left[id_\nu\#N_{e1\nu}^{(1)}(nk_0\#r) -c_\nu\#M_{o1\nu}^{(1)}(nk_0\#r)\right]\,,
\quad r \leq a\,,
\end{equation}
where
\begin{eqnarray}
&&a_\nu= \frac{ n^2\, j_\nu(nk_0a)\, \psi_\nu^{(1)}(k_0a) -  j_\nu(k_0a)\,\psi_\nu^{(1)}(nk_0a)}
{ n^2\, j_\nu(nk_0a)\, \psi_\nu^{(3)}(k_0a) - h_\nu^{(1)}(k_0a)\,\psi_\nu^{(1)}(nk_0a)}\,,
\\
&&b_\nu=\frac{ j_\nu(nk_0a) \,\psi_\nu^{(1)}(k_0a) -   j_\nu(k_0a)\,\psi_\nu^{(1)}(nk_0a)}
{  j_\nu(nk_0a) \,\psi_\nu^{(3)}(k_0a) -  h_\nu^{(1)}(k_0a)\,\psi_\nu^{(1)}(nk_0a)}\,,
\\
&&c_\nu=\frac{ i/(k_0a)}
{  j_\nu(nk_0a) \,\psi_\nu^{(3)}(k_0a) -  h_\nu^{(1)}(k_0a)\,\psi_\nu^{(1)}(nk_0a)}\,,
\\
&&d_\nu= \frac{ in/(k_0a)}
{ n^2\, j_\nu(nk_0a)\, \psi_\nu^{(3)}(k_0a) - h_\nu^{(1)}(k_0a)\,\psi_\nu^{(1)}(nk_0a)}\,;
\label{eq13}
\end{eqnarray}
$j_\nu(\xi)$ and $h_\nu^{(1)}(\xi)$ are the spherical Bessel function
and the spherical Hankel function of the first kind, respectively; and
\begin{equation}
\psi_\nu^{(1)}(\xi) = \frac{d}{d\xi}\left[\xi\,j_\nu(\xi)\right]\,,\qquad
\psi_\nu^{(3)}(\xi) = \frac{d}{d\xi}\left[\xi\,h_\nu^{(1)}(\xi)\right]\,.
\end{equation}
Expressions for the corresponding magnetic field phasors can be derived using the
Faraday equation.

Let  the change $n\rightarrow -n$ be effected in Eqs.~(\ref{eq7})--(\ref{eq13}). Because $j_\nu(-\xi) = (-1)^\nu j_\nu(\xi)$ and $\psi_\nu^{(1)}(-\xi) = (-1)^\nu \psi_\nu^{(1)}(\xi)$, it follows that both $a_\nu$ and $b_\nu$ are not affected, and neither are the scattered electric and magnetic field phasors. Although $c_\nu\rightarrow (-1)^\nu\, c_\nu$ and $d_\nu\rightarrow (-1)^{\nu+1}\, d_\nu$, the internal electric and magnetic field phasors are not affected because
$\#M_{\sigma m\nu}^{(1)}(-\#w)=(-1)^\nu\,\#M_{\sigma m\nu}^{(1)}(\#w)$
and
$\#N_{\sigma m\nu}^{(1)}(-\#w)=(-1)^{\nu+1}\,\#N_{\sigma m\nu}^{(1)}(\#w)$.
 Therefore, again, the fields \emph{everywhere} do not depend on the choice of the refractive index of the active material. We have numerically verified that the same conclusion holds for a dielectric sphere with a perfectly conducting, concentric,
spherical core.

\subsection{More general scattering problems}
The conclusion garnered in Sec.~\ref{sphere} is valid in a far more general situation.
Let all space be divided into three mutually disjoint regions $V_0$, $V_J$, and $V_\eps$
as follows. Whereas both $V_J$ and $V_\eps$ are bounded
in extent, the vacuous region $V_0$ extends to infinity in all directions. The region $V_\eps$ is filled with an active material, whereas the sources of the electromagnetic field reside wholly in $V_J$.

The solution of the frequency--domain  Maxwell curl equations
\emph{everywhere} can be stated as~\cite{LM1993}
\begin{eqnarray}
\nonumber
&&\#E(\#r,\omega)= i\omega\muo\int_{V_J}\,\=G(k_0\#r,k_0\#r')\.\#J_{so}(\#r',\omega)\,d^3\#r'\\
&&\qquad+\,
k_0^2\,
\int_{V_\eps}\, \les n^2\om-1\ris\=G(k_0\#r,k_0\#r')\.\#E(\#r',\omega)\,d^3\#r'\,, \quad
\#r\in V_J\cup V_\eps\cup V_0\,,
\label{eq15}
\end{eqnarray}
where $\muo$ is the free--space permeability, $\=G(k_0\#r,k_0\#r')$ is the dyadic Green function for free space,
and $\#J_{so}(\#r,\omega)$ is the source electric current density phasor. An expression for the magnetic field phasor can be derived  from Eq.~(\ref{eq15}) and  the Faraday equation.
Once again, the electric and the magnetic  field phasors \emph{everywhere}
are not affected by the change $n\rightarrow -n$. The same conclusion is obtained if the active material is
nonhomogeneous, for which case $n^2\om$ must be replaced by $n^2(\#r',\omega)$
in Eq.~(\ref{eq15}).

\subsection{Homogeneously filled waveguides of uniform cross--section}
A commonplace problem in electromagnetism is the propagation of
waves in straight waveguides of uniform cross--section and perfectly
conducting walls. Let us begin with single--conductor waveguides.
An example is the rectangular waveguide
whose walls are formed by the planes $x=0$, $x=a$, $y=0$, and $y=b$.
Let the propagation direction be along the $+z$ axis. The electric field
phasor in this waveguide may be represented as the modal sum
\cite{VB}
\begin{eqnarray}
\nonumber && \#E(\#r,\omega) =
\sum_{_{p\in\mathbb{Z}}}\,\sum_{_{q\in\mathbb{Z}}} \,  A_{pq}^{(TE)}\,\exp(\gamma_{pq}z)\,
\frac{i\omega\muo}{\Delta_{pq}} \les
-\qb\cpa\sqb\ux\right.
\\
\nonumber
&&\hspace{80pt}+\left.
\pa\spa\cqb\uy\ris
\\
\nonumber &&\hspace{15pt}+
\sum_{_{p\in\mathbb{Z}}}\,\sum_{_{q\in\mathbb{Z}}} \,A_{pq}^{(TM)}\,\exp(\gamma_{pq}z) 
\Bigg\{ \frac{\gamma_{pq}}{\Delta_{pq}}
\Bigg[ \pa \cpa\sqb\ux
\\
&&\hspace{60pt}+ \qb\spa\cqb\uy\Bigg]  +\spa\sqb\uz\Bigg\}\,,
\end{eqnarray}
wherein the combination $p=q=0$
is not allowed; $A_{pq}^{(TE)}$ and $B_{pq}^{(TM)}$ are modal coefficients
for transverse--electric and transverse--magnetic modes, respectively; and
$\gamma_{pq}^2
=\Delta_{pq}-k_0^2\,n^2$ with
$ \Delta_{pq}= (p\pi/a)^2 + (q\pi/b)^2$. A similar expression for magnetic
field phasor can be derived using the Faraday equation. The value of the
propagation constant  $\gamma_{pq}$ of the $(pq)^{th}$ mode is chosen to ensure that the integral of the modal
time--averaged Poynting vector over any $xy$ plane is coparallel with $\uz$.
Neither
the electric nor the magnetic  field phasor is affected by the change $n\rightarrow -n$.
The same conclusion holds true for parallel--plate waveguides ($b\rightarrow\infty$)
and circular waveguides~\cite{VB}.

More generally, in a cylindrical coordinate system $(u,v,z)$, let $\#E(\#r,\omega)=
\left[\#e_t(u,v,\omega) + e_z(u,v,\omega)\uz\right]$ $\exp(\gamma z)$
and $\#H(\#r,\omega)=\left[\#h_t(u,v,\omega) + h_z(u,v,\omega)
\uz\right]\exp(\gamma z)$, where $\uz\cdot\#e_t=0$ and $\uz\cdot\#h_t=0$. Then,
the source--free Maxwell curl equations yield the transverse components of the
field phasors in terms of the longitudinal components as~\cite{VB}
\begin{equation}
\left.\begin{array}{l}
\#e_t (u,v,\omega)  =\frac{\gamma}{\gamma^2+k_0^2n^2}
\les({\rm grad}_t e_z-\frac{i\omega\muo}{\gamma}\,
\uz\times\left({\rm grad}_t h_z\right)\ris\\[5pt]
\#h_t  (u,v,\omega) =\frac{\gamma}{\gamma^2+k_0^2n^2}
\les({\rm grad}_t h_z+\frac{i\omega\epso n^2}{\gamma}\,
\uz\times\left({\rm grad}_t e_z\right)\ris\end{array}
\right\}\,,
\end{equation}
where ${\rm grad}_t\equiv {\rm grad}-\uz\left\{\partial/\partial z\right\}$. The longitudinal
field components satisfy the Helmholtz equation as follows:
\begin{equation}
\left(\nabla^2 -\frac{\partial^2}{\partial z^2} + k_0^2\,n^2 +\gamma^2\right)\left\{
\begin{array}{c} e_z(u,v,\omega)\\ h_z(u,v,\omega)\end{array}\right\}
=\left\{
\begin{array}{c} 0\\ 0\end{array}\right\}\,.
\end{equation}
For guided--wave propagation, $\gamma^2$ emerges from a dispersion 
equation derived from enforcing the usual boundary conditions on 
the perfectly conducting walls. Clearly, to obtain
modal solutions for $e_z$ and $h_z$, the refractive index 
is not needed by itself, but its square is needed.
Therefore, the fields are unaffected whether $n=\nplus$ or $n=\nminus$ is chosen.

In a two--conductor waveguide, in addition to transverse--electric and transverse--magnetic
modes, a transverse--electromagnetic mode
can also propagate along the $+z$ axis. This latter type of mode is like a plane wave, and 
its description does require knowledge of the correct refractive index. The number of
transverse--electromagnetic modes increases with the number of conductors \cite{VB}.
As waveguide sections of finite length are used in practice, modal propagation
along the $-z$ and $+z$ directions must be considered simultaneously
\cite{Johnk}. The modal coefficients
of the counterpropagating modes in a waveguide section would
adjust according to the boundary conditions enforced at both ends
of the  section, just as in Sec.~\ref{slab} for a
slab; thus, the correct choice of the refractive index
need not be critical.

\subsection{Flat lenses}
A major motivation for current research on isotropic
dielectric--magnetic materials that display negative phase velocity
(i.e., ${\rm Re}(n)<0$) is their potential to form flat lenses.
Ideally, a flat lens is a slab made of nondissipative and
nondispersive material with $\eps_r = -1$ and relative permeability
$\mu_r = -1$ (i.e., anti--vacuum). When this lens is sandwiched
between vacuous half--spaces, an image of a source placed on one
side of the lens is formed on the other side of lens and another
image is formed within the lens itself. Both propagating and
evanescent modes are brought to a focus at the image points in this
ideal scenario \cite{Rama2005}. A rigorous frequency--domain
analysis reveals that the creation of images is independent of the
choice of the refractive index of the lens material; furthermore,
the focusing properties are seriously compromised if  $\eps_r =
\mu_r \neq -1$ \cite{Ziolkowski_Heyman}. The imaging capabilities
are further compromised by dissipation and/or dispersion, and the
extent to which the slab demonstrates paraxial focusing or
channeling of the illuminating field is independent of whether the
slab's refractive index is chosen to lie in the upper or lower
half--space in the complex plane \cite{Ziolkowski_Heyman}. From
these results, it follows that the nonmagnetic, active dielectric
material of our interest here is unsuitable for flat lenses.


\section{Discussion}

There remains the problem of reflection of a plane wave
from a half--space filled with an active
material~\cite{NS2007,SkaarPRE}. As has been shown by reference to the
time--domain solution~\cite{GML2007}, only one of the two choices of
the refractive index yields the correct frequency--domain
reflectance; that is also the conclusion from an analysis with the Laplace transform
\cite{SkaarPRE}. However, a time--domain solution can never access an entire half--space; thus, the removal of the ambiguity through a comparative study may not be
considered by some researchers to be totally
compelling. Such a comparative study may not be possible for many
active materials anyway, because the fields will continue to grow, invalidate
the assumption of linearity, and eventually counter the assumption of temporal
stationarity of the constitutive properties. Furthermore, whereas when solving
a frequency--domain problem, a half--space
can be considered to be an adequate simplification of a sufficiently thick slab
when the material is dissipative, the process of taking the limit $L\to\infty$ in Sec.~\ref{slab}~---~for an active material \cite{NC2007,NC2007e}~---~has
unphysical consequences in general \cite{SkaarPRE}. This is because the
back face of an active slab can receive a sufficiently strong field
which it can reflect back to reach the front face, 
but a half--space has no back face.

To conclude, we have shown that, for the most common frequency--domain
problems handled by electromagnetics researchers, knowledge of the \emph{correct}
choice of the refractive index of a linear,
homogeneous, isotropic, active, dielectric material is inessential. Either
of the two choices would serve adequately. The major exception we encountered
is the half--space problem, but that problem is significantly unrealistic.


\end{document}